\documentstyle[10pt]{article}
\def \to {\rightarrow}
\def \beq {\begin{equation}}
\def \eeq {\end{equation}}
\def \ba {\begin{eqnarray}}
\def \ea {\end{eqnarray}}
\def \jpsi {J/\psi}
\def \< {\left <}
\def \> {\right >}

\begin{document}

\title{Production of $0^{++} $glueball from double diffractive process in high energy
        $p+p(\bar p)$ collision}
\author{Hong-An Peng$^{a,b}$,~~Jia-Sheng Xu$^a$ \\
 $^a${\small Department of Physics, Peking University, 
            Beijing 100871, People's Republic of China}  \\
 $^b${\small CCAST (World Laboratory), Beijing 100080, People's Republic of China} }
\maketitle

\begin{abstract}
Motivated by the recent experimental data about candidates for glueball from
different processes, we  discuss in this paper the production of
$0^{++}$ glueball from double diffractive scattering at momentum transfer
$|t|\lower4pt\hbox{$\buildrel <\over\sim$} 1GeV^2$ in high energy $p+p(\bar p)$ collision.
We apply the phenomenology of Pomeron ($I\!\! P$) of
Donnachie-Landshoff, the field theory model of $I\! P$ of Landshoff-Nachtmann
and the relevant calculating approachs. We assume while $I\! P$ coupling
with glueball, the $0^{++}$ glueball can be considered as a bound state
of two non-perturbative massive gluons.
We  evaluate the dependence of cross section for $0^{++}$ glueball
production on system energy $\sqrt{s}$ and show that it could be tested
experimentally.
\end{abstract}
~~~~~~~~~~
\vskip 1cm
PACS number(s): 12.40Nn, 12.39.Mk, 13.85.Lg     \\

\vfill\eject\pagestyle{plain}\setcounter{page}{1}

\section{Introduction}

Since the color $SU(3)$ gauge theory possess of self-coupling characteristic
between gluon fields, it had predicted long before the quantum
chromodynamics (QCD) became the basic dynamics of strong interaction,
that there would  exist states in hadronic spectroscope which are formed
from purely gluon field---glueball\cite{1p}. Though now there are still
lack of strict theoretical proving and conclusive experimental evidence
to confirm its existence,  in recent couple of years there are a lot
of reports from relevant data analysis that several candidates of glueball
and/or states  mixing with quark pair have been observed experimentally
\cite{2p}.
It is well believed that glueballs
should easily be produced in processes with rich source of gluon constituent,
thus experimentalists concentrate their attention to seek glueballs on
$p\bar p$ annihilation \cite{3p}, $\jpsi$ radiative decay\cite{4p}, and
high energy central $h - h$  collisions\cite{5p}.
Glueball production from $\jpsi$ radiative decay has been analyzed in a model
independent manner by Close {\it et.al.}\cite{6p}, but yet there has not
been any analysis in central $h-h$ collision and in
$p\bar p$
annihilation.

\par
At high energy central $h - h $ collisions the increase in cross section
with increased center of mass energy \cite{arms} is consistent with the
double pomeron exchange (DPE) machanism , which was predicted to be a
source of gluonic states\cite{robson}.
A large glueball production cross section
in the central region is predicted by Gershtein and Logunov, they related
the rise of total cross section with increasing energy to the exchange
of glueballs in the $t- $channel or to the collising of the soft gluon
seas of the interacting particles \cite{gerl}.
These observations tell us that glueball production in high energy h - h 
central collisions would
be intimately connected with the non-perturbative mechanism .

\par
In this paper, we discuss
the glueball production from double diffractive scattering in
high energy $p + p ({\bar p} ) $ collisions,
\beq
p+p(\bar p) \to p+p(\bar p) +G,
\eeq
as a first attempt, we consider $J^{PC}=0^{++}$ glueball only.

\par
Our approach is on the basis of the phenomenology
of $I\! P$ from Donnachie-Landshoff\cite{dl}, the field theory model of
$I\! P$ from Landshoff-Nachtmann\cite{ln}.
Based on the above approach,
a more refined and sophisticated  model of diffractive scattering
had proposed by
Cudell and Hernandez \cite{c96}, concentrating on Higgs diffractive
production in pp collision $ p + p \to p + p + H $, they obtained a result
close to that of Bialas and Landshoff\cite{bl} using the L-N model of DPE.

\par

Since the glueballs are produced through  DPE, it could
believed that they
are laid predominately in the central region of final state rapidity
distribution and with large symmerty rapidity gaps from each of the final
diffractive protons ( or antiprotons) which are the characteristic properties
of DPE processed as Eq.(1) . In principle, it should be easily detected and
according to it we could distinguish it from all other strong backgrounds.
We will come back to this  point in last section.

\par
Of how to link L-N field theory of $I\!P$ with the phenomenology of
Regge behaviour, we follow the approach of Bialas and Landshoff\cite{bl}.
Since the information of coulping of $0^{++}$ glueball with nonperturbative
gluon  coupling is very scarely , we
adopt the color singlet approximation \cite{berg}, which always used in heavy
qurak physics. Under this approximation  the coupling
structure is $ B g^{\alpha \beta }
\delta _{eh} $ , where $\alpha ~ e ~ etc. $ are the Lorentz and color
indexes of the two  non-perturbative gluons respectively , the  parameter B
can be estimated from
the branching ratio of $ \jpsi $ radiative decay to $0^{++}$ glueball.

\par
In the following we describe in detail our calculating scheme in Sec.II,
we give the formalism and derive the coupling constant $B$ of
the two nonperturbative  gluons with $0^{++} $ glueball  in
Sec.III, our results and some related discussions are given
in last Sec.IV.

\section{Calculating Scheme}

In phenomenology of $I\! P$ given by  Donnachie-Landshoff\cite{dl},
the nucleon is to be treated as a bound state composed of
three clusters, they are  formed from non-perturbative QCD
effects and their size are much smaller than nucleon radius. The core of
each cluster is a valence quark, with sea quarks and gluons surrounded.
Furthermore D-L have demonstrated that in coupling with nucleon,
to a good approximation,the
$I\!P$ behaves like a isoscalar ( C= + 1) photon.
So it means that the coupling of $I\! P$ with a nucleon is
actually that of $I \! P$ with the three clusters (constitude quark ) in
an incoherent manner. In this picture $I \! P$ model one does not need to consider
the interaction of $I\! P$ with any other parton (sea quark or gluons) in
the nucleon. With this model Donnachie and Landshoff have successfully
explained a lot of experimental effects in high energy soft strong
interaction processes\cite{l92}.

In discussing glueball production from DPE , we adopt the same
point of view: we should not start such a problem from parton model 
and perturbative QCD, but instead from non-perturbative QCD at all and assume
the glueball to be a bound state of two small
flavorless clusters (or ``constituent'' gluons) which  are formed from
gluons by non-perturbative QCD effects, and are massive.
Like the nucleon case, we assume the coupling of $I\! P$ with glueball
is actually
that of $I\! P$ with these ``constituent'' gluons.
Also, we do not need  to consider the interaction of $I\! P$ with any other
massless gluons in the glueball.

\par
From the status  both of the available data for elastic diffractive
scattering and the effectuality of D-L $I\!P $ model in high energy
h-h collisions, we restricted in this paper the momantum transfer of
scattered proton (antiproton) to  to be small ,${\sl{ e.g }}$ 
$|t_i| \lower4pt\hbox{$\buildrel <\over\sim$} 1.0 GeV^2 $ .
The whole picture is  sketched in Fig.1.

\par
From Regge pole theory\cite{cls} and $I\! P$ model of D-L, when $s \ll s_1$,
$s_2\ll m_N^2$, and $|t_i|\lower4pt\hbox{$\buildrel <\over\sim $} 1GeV^2$, the asymptotic form of the amplitude for
Fig.1 is
\beq
\label{bmatr}
(\frac{s}{s_2})^{\alpha _{I\kern-.20emP}(t_1)-1}(\frac{s}{s_1})^{\alpha _{I\kern-.20emP}
(t_2)-1}
F^{(G)}(\alpha _{I\kern-.20emP}(t_1) , \alpha _{I\kern-.20emP}(t_2) , z)
 F_{1}(t_1)F_{1}(t_2) \gamma _{\lambda }\otimes
 \gamma ^{\lambda }
\eeq
where $t_i=(p_i-p_i^\prime)^2$, $z=\frac{s_1s_2}{sm^2}$, and $m$ is the
$0^{++}$ glueball mass,
$\alpha_{I\! P}(t)$ is Regge pole trajectory of $I\! P$.
From data analysis \cite{dl} when $|t|$ is
small, $\alpha_{I\! P}(t)\approx \alpha_{I\! P}(0)+\alpha_{I\! P}^\prime t=
1.086+0.25GeV^{-2} t$ .$ F_{1}(t)$ is the Dirac form factor of proton.
The direct product of gamma
matrix shows that the external proton lines will be traced when
calculate the cross section.

\par
Since the diffractive scattering condition requires
\beq
\frac{s_1}{s},~\frac{s_2}{s}<\delta
\eeq
and usually assume $\delta = 0.1$, hence in  the asymptotic form Eq.(2)
one could neglect
lower powers of $({s\over s_1})$ and $({s\over s_2})$.
$F^{(G)}(\alpha_{I\! P}(t_1),\alpha_{I\! P}(t_2),z)$
is the
$I\! P-I\! P-G$ vertex function, it general structure is known\cite{drum},
but its concrete form must be further considered.

Since $I\!P$-nucleon  coupling should be considered as pomeron
incoherently coupled with three constituent quarks, and according to field
theory model, the $I\! P$ exchange is just the exchange of two
non-perturbative gluons\cite{ln}\cite{low}, thus Fig.1 should be concreted
as Fig.2.

\par
In the field theory model of pomeron, the $I\!P - I\!P - G $ vertex function
$F^{(G)}(\alpha_{I\! P}(t_1),\alpha_{I\! P}(t_2),z)$ for
$|t_i| \lower4pt\hbox{$\buildrel <\over\sim$} 1.0 GeV^2 $
can be calculated at
by the sum of Feynman amplitudes of Fig.2 which as
shown in \cite{bl}  is equal
to its $s-$channel discontinuity of the first diagram, thus as Fig.3.
This simplify our calculation greatly. The black blob in Fig.3 is the vertex
function of $0^{++}$ glueball with gluons of $I\! P$ which will be discussed
in next section.

\par
In the following we first get the forward diffractive scattering amplitude
for Eq.(1) in L-N field theory model of $I\!P$, then identify it
with the corresponding Regge behaviour formula Eq.(2) at $t_i =0 $ to obtain
the normalized value of $I\!P - I\!P - $ glueball vertex function 
$F^{(G)}(\alpha_{I\! P}(t_1),\alpha_{I\! P}(t_2),z) |_{t_1 = t_2 = 0 }$ .
\footnote{Since in D-L model,
 $\alpha_{I\! P}(t)|_{t=0} \equiv 1 + \varepsilon \approx 1.086$,
so it seems having some inconsistent with this identification, which
has explained in \cite{bl}. Furthermore, this problem has studied
by Ross \cite{ross}
who using a hybrid model in which one uses PQCD to treat
interactions  between gluons, but takes a D-L type non-perturbative
propagator for the gluon.
In the leading logarithm approximation, by summing a dominant subset of
diagrams, the phenomenolog required Regge behaviour  is  obtained. }
After this has done  we continue  Eq.(2)  to $|t_i| \ne 0 $ and discuss DPE
production of glueball.

\section{Formalism and Coupling Constant $B$}

Using the Sudakov variables to calculate the loop integral of Fig.3 is very
convenient, here
\ba
\label{suda}
k& = &\frac{\bar x p_1}{s} + \frac{\bar y p_2}{s} + v  \nonumber \\
p^{\prime}_{1}& = & x_1 p_1 + \frac{\bar y_{1}p_2}{s} + v_{1}  \nonumber \\
p^{\prime}_{2}& =& \frac{\bar x_{2}p_1}{s} + y_2p_2 + v_{2} 
\ea
We have put the light quark mass equal zero, the $v$, $v_1$, and $v_2$ are
transverse to $p_1$ and $p_2$ and so effectually are two dimensional,
$v^2 = -{\bf {v} ^2},v_{i}^{2} = - {\bf {v} }_{i}^{2}$. Then

\ba
\label{t1t2}
t_1&=&(p_1-p_{1}^{\prime })^2=-{\bf {v}}^{2}_{1}/x_{1} \nonumber \\ 
t_2&=&(p_2-p_{2}^{\prime })^2=-{\bf {v}}^{2}_{2}/y_2 \nonumber \\ 
s_1& \sim & s(1-y_2) \nonumber \\
s_2& \sim & s(1-x_1) 
\ea

As mentioned above, we first need the $s-$ channel discontinuity of Fig.3
at $t_1 \approx t_2\approx 0$, so we set $v_1\approx v_2\approx 0$, then

\ba
\label{sudaint}
\int d^{4}k{\delta }(q^{2}_{1})\delta (q^{2}_{2}) & \sim  &
         \frac{1}{2s}\int d{\bar x} d{\bar y}d^{2}v\delta ({\bar y} -{\bf v}^{2})
         \delta (-{\bar x}-{\bf v}^{2}) ,   \nonumber \\
\int d^{4}p_{1}^{\prime}\delta ({p_{1}^{\prime }}^{2}) & \sim &
    \frac{1}{2}  \int dx_1 d{\bar y}_1 \delta (x_1 {\bar y}_1)d^2{\bf v}_{1} ,
    \nonumber \\
\int d^{4}p_{2}^{\prime}\delta ({p_{2}^{\prime }}^{2}) & \sim  &
    \frac{1}{2}  \int d{\bar x}_2 d y_2 \delta ({\bar x}_2  y_2)d^2 {\bf v}_{2} .
\ea

Here we explain how we can get from Fig.3 the direct product from
$\gamma_\lambda\otimes\gamma^\lambda$ as in Eq.(2). The upper quark
line in Fig.3 has gamma matrices $\gamma^\mu\gamma\cdot q_1\gamma_\lambda$,
for large $ s $ its asymptotic form equivalent to
\beq
2q_1^\mu\gamma^\lambda,
\eeq
because in calculating the differential cross section, we also need to
multiply $\gamma\cdot p_1$ on the left and $\gamma\cdot p_1^\prime$
on the right of this expression.
When these are included, the difference between (7) and the original
expressions gives contribution to cross section  are of order $\delta$.
Similarly, from the lower quark line we obtain $2q_2^\nu\gamma_\lambda$.

\par
Now from Fig.3, the amplitude of $q-q(\bar q)$ diffractive scattering through
DPE at $t_1=t_2=0$ is
\ba
\nonumber
M^{(G)}&& \gamma ^{\lambda }\otimes \gamma ^{\lambda },\\
\label{jqwg4}
M^{(G)}&&=\frac{ig^{6}}{2{\pi }^{2}s} 
        \int d^{2}v W^{(G)}D(-{\bf v}^{2})
         D(-x_1 {\bf v}^{2})D(-y_2 {\bf v}^{2}) ,    
\ea

where $D(q^2)$ is non-perturbative gluon propagator of $I\! P$,
$g$ is the coupling constant between non-perturbative gluon with quark,

\beq
g^2 D(q^2)= - A~exp(\frac{q^2}{\mu ^2}),
\eeq
where $\mu = 1.1 GeV$, $A^2\mu^2=72 \pi \beta^2$ ,$\beta ^2 = 3.93
 GeV^{-2} $ is
$I\!P-$nucleon coupling constant in the pomeron D-L model\cite{cudl90}.

Using the approximation matrix form Eq.(7), we found from Fig.3, the $W^{(G)}$
in Eq.(8) ,
\beq
W^{(G)}=q_1^\mu A_{\mu\nu}^{(G)}q_2^\nu,
\eeq
where vertex function $A_{\mu\nu}^{(G)}$ connecting the gluons of $I\! P$
with $0^{++}$ glueball. The simplest diagrams for this vertex function shown
in Fig.4 and Fig.5.

\par
In deriving these vertex functions, for calculating simplicity but not
necessary we have used color singlet model approximation\cite{berg}
for the vertex, which require gluon lines cutting by vertical dashed line in
these figures are limited on mass shell. Thus from Fig.4 we get

\ba
\label{amug3g}
A^{G_3}_{\mu \nu,bc }&=&2 B F_{\mu \alpha \lambda }(k_1, q , -\frac{1}{2} P )
   D(q^2)
      g^{\alpha \beta }F_{\beta \nu \rho }(-\frac{1}{2} P , k_2 ,-q)
      g^{\rho \lambda } \cdot f_{beg}f_{hcg}\cdot \delta_{eh}  ,
\ea
where
\ba
F_{\mu _ 1 \mu _2 \mu _3 }(r_1 ,r_2,r_3 )& = &
 (r_1 -r_2)_{\mu _3}g_{\mu _1 \mu _2} +(r_2-r_3)_{\mu _1}g_{\mu _2 \mu _3}
 +(r_3 -r_1)_{\mu _2}g_{\mu _3 \mu _1}  \nonumber  \\
 r_1+r_2+r_3& = & 0.
\ea
From Fig.5 we get
\beq
A_{\mu\nu,bc}^{G_4} =  -12  B g_{\mu\nu} f_{beg}f_{hcg} \delta_{eh}.
\eeq

From Eqs.(10) to (13) we get
\ba
W^{G_3}& = &\frac{- B~D(q^2) }{4 s^2 }
            \cdot(5 m^2 s^3 + 10 m^2 s^2 {\bf v}^2 + 16 m^2 s {\bf v}^4
     + 15 s^3 {\bf v}^2 x_1+ 15 s^3 {\bf v}^2 y_2 - 10 s^3 {\bf v}^2
    \nonumber \\
& &     + 6 s^2 {\bf v}^4 x_1
        + 6 s^2 {\bf v}^4 y_2 + 4 s^2 {\bf v}^4 + 12 s {\bf v}^6 x_1
        + 12 s {\bf v}^6 y_2 + 24 {\bf v}^8)  \nonumber \\
q^2 & = & \frac{-1}{4s} (m^2 s  + 2 s {\bf v}^2 x_1 + 2 s {\bf v}^2 y_2 +
         4{\bf v}^4) \nonumber \\
W^{G_4} & = & - 12 B (\frac{1}{2} s +  \frac{ {\bf v}^4 }{s} ) \nonumber\\
W^{(G)} & = & W^{G_3} + W^{G_4}
\ea

Form Eqs.(8)(9)(14) , we get $M^{(G)}$,
for Fig.3, which is
$
F^{(G)}(\alpha _{I\kern-.20emP}(t_1) , \alpha _{I\kern-.20emP}(t_2) , z)
|_{t_1 = t_2 = 0 } $ as we explained in section II .

\par
For $t_1,~t_2\not = 0$, the general structure of $F(\alpha_{I\! P}(t_1),
\alpha_{I\! P}(t_2),z)$ can be analyzed from Regge theory and this has been done
by Drummon {\it et. al.}\cite{drum}. They have shown that when $|t_1|$, $|t_2|$
are small,
 $F(\alpha_{I\! P}(t_1),\alpha_{I\! P}(t_2),z)$
  are finite for any
$z$. This point is very important for us since the D-L $I\! P$ model,
in strictly speaking, is valid only for $|t_1|,~|t_2| \lower4pt\hbox{$\buildrel <\over\sim$} 1GeV^2$.
But we can also see from Eqs.(2) and (3), all contributions from large
$|t_1|,~|t_2|$ are higher powers of $\delta$, and they can be neglected safely.
So, for $t_1,~t_2\not = 0$,we have
\beq
 F(\alpha_{I\! P}(t_1),\alpha_{I\! P}(t_2),z)|_
 {|\!t_1\!|,|\!t_2\!| \leq 1 GeV^2 }
 \simeq
 F(\alpha_{I\! P}(t_1),\alpha_{I\! P}(t_2),z) |_{t_1 = t_2 =0 }
 = M^{(G)}
\eeq
 Thus the amplitude $T^{(G)}$ for glueball production
in $p-p(\bar p)$ high energy double diffractive scattering from D-L and L-N
$I\! P$ model and formalism, in a good approximation, can be got from
Eqs.(2) and (15) and is

\beq
T^{(G)}=9 M^{(G)}({\frac{s}{s_2}})^{\alpha _{I\kern-.20emP}(t_1)-1}
        ({\frac{s}{s_1}})^{\alpha _{I\kern-.20emP}(t_2)-1} F_1(t_1)F_1(t_2)\gamma _{
        \lambda }\otimes \gamma ^{\lambda } .
\eeq
where factor $9$ comes from $3$ quarks in nucleon, $F_1(t)$ is the Dirac
form factor of proton, using the exponential approximation for $F_1(t) $
at small $ |t| $ : $F_1(t) \simeq e^{\lambda t}$, $\lambda \simeq 2 GeV^2
$, the cross section formula is
\ba
\label{sgg4}
\sigma ^{(G)}&=&\frac{F_{c}^{G}}{2(2\pi )^5} (\frac{s}{m^{2}})^{2\varepsilon }
          \int \frac{dx_1}{x_1} \frac{dy_2}{y_2}{\mid M^{(G)}\mid}^2 \delta {\big (}(1-x_1)(1-y_2)
          -\frac{m^{2}}{s}{\big )} \nonumber \\
    & & \cdot \int d^2 v_1 d^2 v_2 (1-x_1)^{2 \alpha ^{\prime }{\bf v}^
    {2}_{1}/x_1} 
     (1-y_2)^{2 \alpha ^{\prime }{\bf v}^{2}_{2}/y_2}
     \cdot e^{-2\lambda ({\bf v}^{2}_{1}/x_1+{\bf v}^{2}_{2}/y_2)}
\ea
where $F_c^G=\frac{4}{9}$ is the color factor.
\par
Let us now consider the coupling constant $B$ of constituent gluons with
$0^{++}$ glueball, which could be fixed by radiative decay mode
$\jpsi\to \gamma +f_0(1500) $ as shown in Fig.6, 
\footnote{Close argue in \cite{14p}
that the scalar $f_0(1500)$ may be a glueball$-q\bar q$ mixture, here for
simplicity we assume it is a pure glueball.} 
where
\ba
p_i & =& \frac{1}{2} (p_J-p_G) \nonumber \\
p_j & =&k-\frac{1}{2}p_J \nonumber \\
p_k & =&\frac{1}{2}(p_G-p_J)  \nonumber \\
g_1 & =& g_2=\frac{1}{2} p_G
\ea

\par
It is well know that the $\jpsi $ is dominatly an S- wave state and the
non - relativistic quark potential mode is very successful in describing
the static properties of the $\jpsi $. Inclusive quarkonium decays into
light hadrons are accessible to PQCD\cite{sl97} . For heavy-quarkonium, the
annihilation time $\sim \frac{1}{m_Q} $ is much smaller than the time
scales relevent to $Q{\bar Q} $ binding. Hence the short - distance
annihilation of the $ Q {\bar Q} $ pair can be separated from the long-
distance effect of $Q{\bar Q} $ binding. In our
$\jpsi \to \gamma + G $ case, the long- distance effects of $Q{\bar Q}$
and $g g$ binding  can be separated, the $g g $ binding
effect is included in the non-perturbative coupling constant B of
constituent gluons with the $0^{++} $  glueball .

\par
Using the color singlet model approximation\cite{berg} in $\jpsi$
vertex as usual, we get the amplitude of $\jpsi \to \gamma + f_0$
\beq
\label{mj4x}
 M^{J}= 2A_0 e_q g_G^2 B(M^{(a)}_{\mu \nu \rho } + M^{(b)}_{\mu \nu \rho }
        + M^{(c)}_{\mu \nu \rho } ) g^{\rho \nu }
         \varepsilon _{\mu }(\gamma )
\eeq
where $\epsilon^\mu (k)$ is the polarization vector of photon, $e_q={2\over 3}e$
is the charge of charm quark, coefficient $A_0$ fixed from $\Gamma_{J\to e^+e^-}$:
\beq
A_0^2 e_q^2 = \frac{\Gamma _{J\to e^+e^- } m_J }{2\alpha }
\eeq
and
\ba
\label{mabc4}
M^{(a)}_{\mu \nu \rho }&=&\frac{1}{\sqrt 2}Tr[\gamma \cdot \epsilon _J
  (\frac{1}
  {2} \gamma \cdot p_J +m_c)\gamma _{\mu }\frac{\gamma \cdot p_j+m_c}{p_j^2
  -m_c^2}\gamma _{\nu }\frac{\gamma \cdot p_i+m_c}{p_i^2-m_c^2}
  \gamma _{\rho } ] \nonumber \\
M^{(b)}_{\mu \nu \rho }&=&\frac{1}{\sqrt 2}Tr[\gamma \cdot \epsilon _J
  (\frac{1}
  {2} \gamma \cdot p_J +m_c)\gamma _{\nu }\frac{\gamma \cdot p_k+m_c}{p_k^2
  -m_c^2}\gamma _{\rho }\frac{-\gamma \cdot p_j+m_c}{p_j^2-m_c^2}
  \gamma _{\mu } ] \nonumber \\
M^{(c)}_{\mu \nu \rho }&=&\frac{1}{\sqrt 2}Tr[\gamma \cdot \epsilon _J
  (\frac{1}
  {2} \gamma \cdot p_J +m_c)\gamma _{\nu }\frac{\gamma \cdot p_k+m_c}{p_k^2
  -m_c^2}\gamma _{\mu }\frac{\gamma \cdot p_i+m_c}{p_i^2-m_c^2}
  \gamma _{\rho } ]
\ea

Put Eqs.(18) and (19) into (17), after sum and average with initial states
and final states respectively, we get
\beq
\sum^{--}|M^{J}|^2 =\frac{2}{3}A_0^2 B^2 e_q^2 g_G^4  
\frac{1024 (m^6-m_j^2 m^4 + 11 m^2m_j^4 +m_J^6 )}
             {m_J^4 (m_J^2-m^2)^2},
\eeq
then decay width of $\jpsi \to \gamma + f_0$ is
\beq
\Gamma _{J\to \gamma +f_0} =\frac{m_J^2-m^2}{16 \pi m_J^3} F_c^J
     \sum^{--}|M^J|^2,
\eeq
where color factor $F_c^J={2\over 3}$.

Put Eqs.(18), (20), and (21) together, we get
\beq
B^2 =\frac{\Gamma _{J\to \gamma + f_0}}{\Gamma _{J\to e^+e^-}}
 \frac{9\pi \alpha m_J^6 (m_J^2 -m^2)}  {128 g_G^4 (m_J^6 +
 11 m_J^4 m^2 - m_J^2 m^4 +m^6)}.
\eeq
Decay width $\Gamma_{J\to e^+e^-}$ and $\Gamma_{J\to \gamma +f_0}$ can
be found in \cite{15p},$\frac{\Gamma _{J \to \gamma + f_0 }}{\Gamma _{
J \to e^+e^-}} = \frac{8.2 \times 10^{-4}}{6.02 \times 10^{-2}} $
so as $g_G$ and $m$ are fixed, one gets coupling
constant $B$. For $\frac{g_G^2}{4 \pi } = \frac{12 \pi }{50 ln[m_J / \Lambda
] }, \Lambda = 0.20 GeV , m = 1.5 GeV $ , we get$ B^2 = 3.8 \times 10^{-6}
GeV^2 $ .

\section{Results and Discussions}

We first give a brief comments on the parameters used in our paper.
\begin{enumerate}
\item $0^{++}$ glueball mass $m$ .
\par
  From several models ({\it e.g.} lattice QCD, bag model, potential model,
  and QCD sum rule) analysis, the lowest state mass of $0^{++}$ all are
  fixed at $1.5\sim 1.7GeV$\cite{14p}. Experimentally, relevant data also
  manifest a clear signal of $0^{++}$ resonance state at about
  $1.5GeV$ in the
  central region of high energy $p-p$ collision, thus we let $m=1.5GeV$.

\item Nonperturbative coupling constant  $g$.   
 \par
 In the Abelian gluon theory of Landshoff and Nachtmann, the nonperturbative
 gluon only couples to quark, thus $ g^2 D(q^2) $ always appear together
 and can be  normalised by the constant A in Eq.(9) , the nonperturbative
 constant does not enter the calculation of LN model.
 \footnote{ Even in process which are absent for gluon selfinteractions, may
  meet the same troubles. For example, in discussiong the diffractive
 Higgs production process $p + p \to p + p + H $, only after supposing
 implicitely that coupling constant between top quark with $I\!P$  equals
 to that of u, d quark with $I\!P$, then constant g can be all absorbed
 in Eq.(9). }
 But in non-Abelian  case , especially including gluon self-interactions,
 as showed in Fig.4 and Fig.5, g connot be all absorbed by normalized
 condition Eq.(9), so 
 g enters indeed into our the calculation. Unfortunately our knowledge for the
 non-perturbative coupling constant
 $g$ is very poor now ,  in order to get a sensible answer for
 the gluon structure function at small $x $ \cite{cudell}
 $\alpha _n =\frac{g^2}{4 \pi }$ of order one  is needed, here we take the
 value  $\alpha _n = \frac{g^2}{4\pi } \sim 0.7 $.
\end{enumerate}

Putting these parameters into Eq.\ref{sgg4}, the double diffractive
production cross section of Eq.(1) for
$\sqrt{s}$ from $20 $GeV to $2 \times 10^4 $ GeV, where
$|t_1|,~|t_2|\lower4pt\hbox{$\buildrel <\over\sim$}1GeV^2,
\delta = 0.1 $
are evaluated, as shown in Fig.7.
We see for $\sqrt{s}= 23.7 GeV , 29 Gev , 630 GeV (SppS)$ and 1.8 TeV
(Tevatron) energies, the double diffractive production
cross sections 
are $1.6 \times 10^2 nb, 2.5 \times 10^2 nb ,2.8{\mu \!b}$ and
$4.6 {\mu \!b}$ respectively.
In  the Joint CERN-IHEP experiment in $ 300 GeV $ central $\pi ^- N $ collisions
($\sqrt{s}= 23.7 GeV$ ),the production cross section in the range
 $ 0 < x_F < 0.3 $ (the experiment setup lead to the acceptance is zero
 for $x_F < 0 $, ) is $ 0.2 \pm 0.1 \mu b $ .
Taking account of the addtive quark rule, this corresponds to
 $0.4  - 1.4 \mu b $
 in $P-P({\bar P })$ central collisions at the same center-of-mass
 energy in the range $ -0.3 < x_F < 0.3 $ .
Since $f(1500) $ is produced dominately at small $|t| $
\cite{arms}\cite{wa91}, our results are reasonble .

\par
We have restricted ourseleves to calculate elastic diffractive production
process. It is easily extend to diffractive dissociation processes by
removing the form factor $F_1 (t_i) ( i = 1,2 ) $ in Eq.(2) and thereafter
, the resulting cross sections are also show in Fig.7. We can see that
as center of mass energies  increased the ratio of the cross section of
diffractive dissociation to that of elastic one is reduced from 6 at
$\sqrt{s} = 23 GeV $ to 3 at $\sqrt{s} = 20 Tev $ .

\par
Throughout the calculation we work in Feynman gauge. If one were to work in
another gauge  the
function $D(q^2)$ must be substantially modified to ensure that physical
observables are gauge independent .

\par

In connection with the problem of experimentally detecting the process
Eq.(1), we make following discussions:
\begin{enumerate}
\item 
Since glueballs in process Eq.(1) are produced through DPE in high energy
$p-p({\bar p})$ collision, they should be laid predominately in the
central region of rapidity distribution of final particles and  with  large
symmetry rapidity gaps from  each  of the final proton
( antiproton ) direction. However several effects would be weaken the
gap rate. First, althrogh the glueball and its decay product are colorless
, as final state interacting there are get having survival soft color
interactions between pre-glueball with outgoing pro-nucleons .
The estimated gap survival rate to be $10^{-2} $ at Fermilab
Tevatron \cite{bjk}. Secondly, the produced glueball is a light object ($m_G \sim 1.5
GeV $), they will be produced in
a broad rapidity range, for example, at Tevatron those glueballs
are in the rapidity region $-5.9 <  \eta < 5.9 $ .

\item 
We take the D-L $I\!P$ model and DPE process to discuss glueball production
process at small $|t| $ , like all other high energy diffractive processed
discussed by using the same model, the common remarkable character and
parameter independent property are the energy dependence of total cross
sections which are proportional to
$s^{2 (\alpha_{I\!P} (0) - 1 )}$ . So the cross section in this model
always increases slowly with the increase of the center of mass energy,
we could use
this point to distinguish this production mechanism of glueball from
others, especially from gluon - gluon fusion mechanism in parton model,
since for later the production cross section would be decreased rapidly
as energy increases, perhaps this is the most effective way to test
our results if one could measure the total cross section of process
Eq.(1) over  a large energy range.
\end{enumerate}

\par
In conclusion, using the field theroy model of pomeron exchange and the color
singlet approximation of glueball, we obtain a parameter-free prediction
of the cross section of $0^{++} $ glueball in diffractive production ,
which combineing with the
experiment measureing of $ \sigma (h h \to h h G ) \cdot BR (G \to h1 +
h2 + \cdots) $
can output the important quantities $BR ( G \to h1 + h2 + \cdots )$ , saying
$ BR ( G \to K {\bar K } )$, or if we
 knew $ BR (G \to h1 + h2 + \cdots  )  $ 
then glueball production would be a test of the approach we used.

\vskip 1cm
\begin{center}
{\bf\large Acknowledgments}
\end{center}

This work is supported in part by the National Natural Science Foundation of
China, Doctoral Program Foundation of Institution of Higher Education of China
and Hebei Natural Province Science Foundation, China.

We thank Professors Yu-Ping Kuang, Zhi-Yong Chao, Zhen-Ping Li, and Dr. Feng Yuan for
useful discussions.

%\newpage

\newpage
\centerline{\bf \large Figure Captions}
\vskip 2cm
\noindent
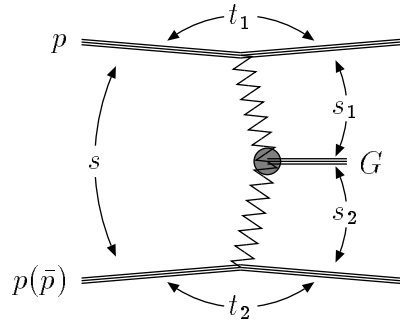
Fig.1. Sketch diagram for Eq.(1) via DPE.

\noindent
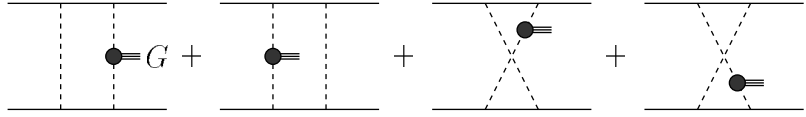
Fig.2. $G$ production in $q-q(\bar q)$ scattering via DPE, dotted lines are
non-perturbative gluon of $I\! P$.

\noindent
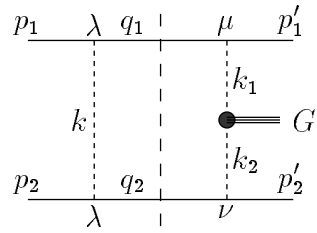
Fig.3. $s-$channel discontinuity of the first diagram in Fig.2.
$q_1$, $q_2$ lines are limited on their mass shell.

\noindent
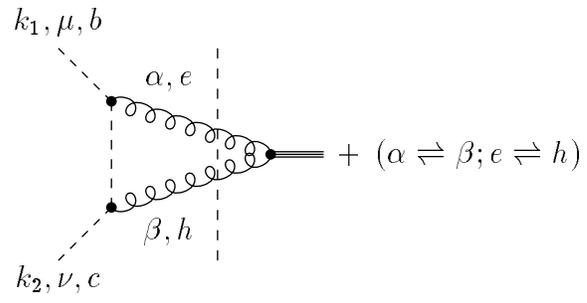
Fig.4. Vertex of glueball and gluons of $I\! P$. Vertical dashed line limit
cutting lines on mass shell.

\noindent
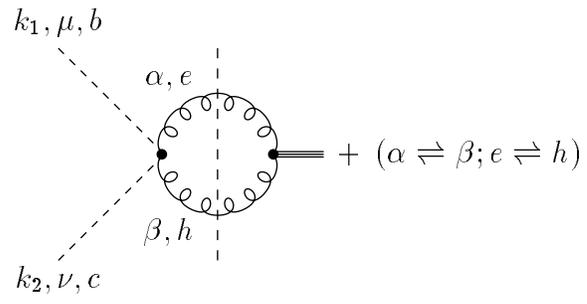
Fig.5. The same as Fig.4, but for four gluons vertex.

\noindent
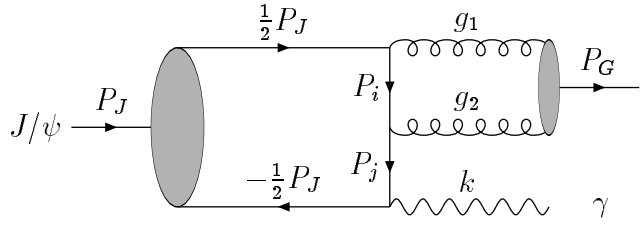
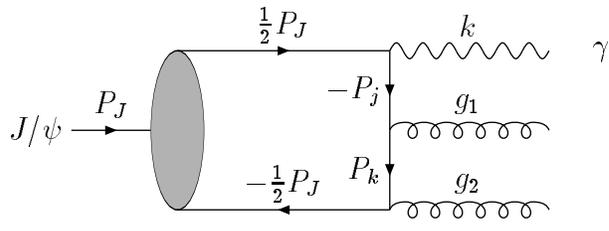
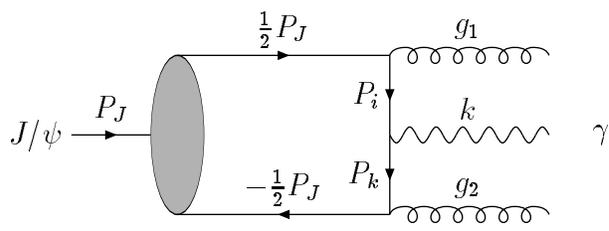
Fig.6. Glueball production in radiative decay of $J/\psi$.

\noindent
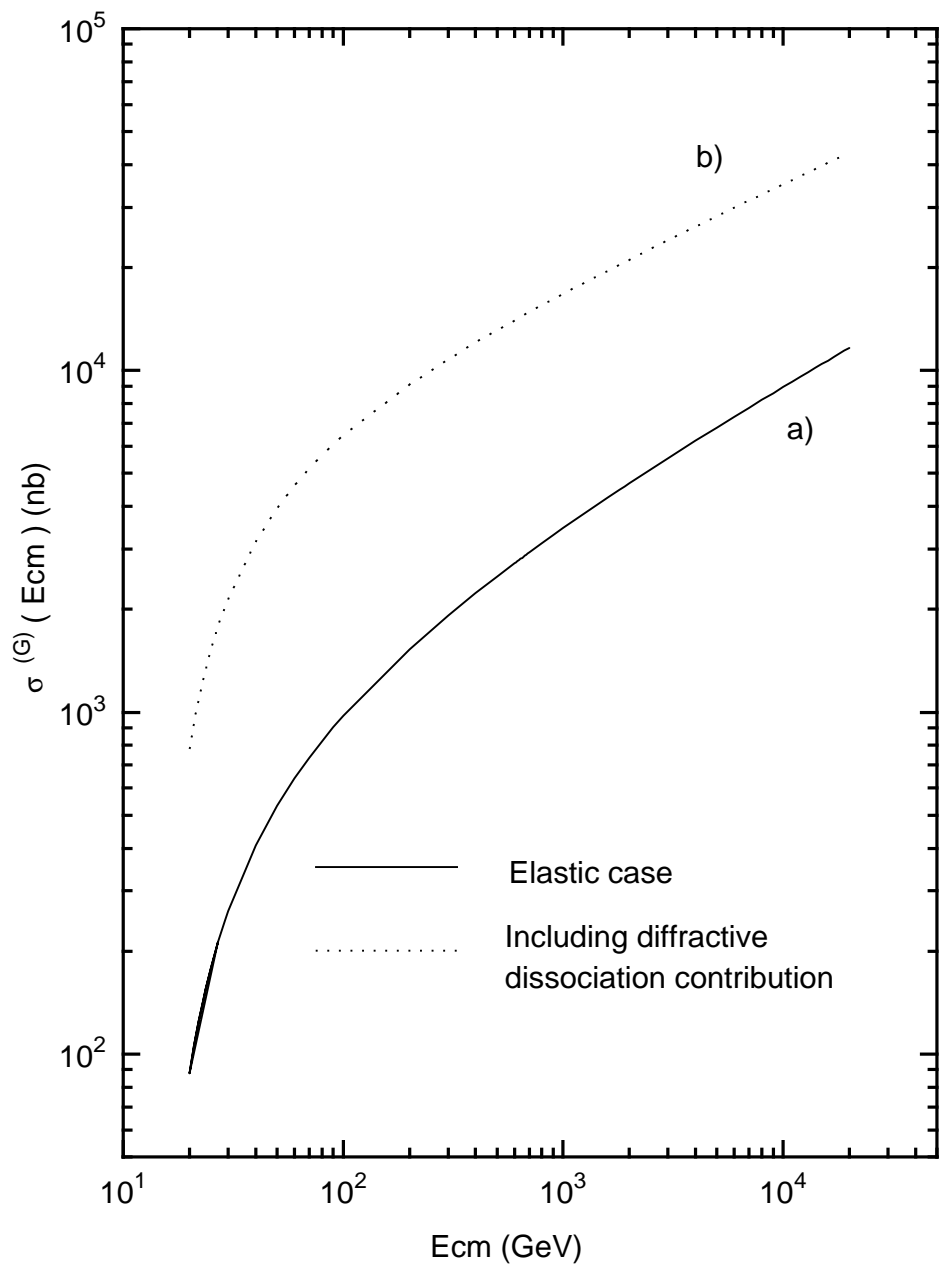
Fig.7. Production cross section of $0^{++}$ glueball  in
double diffractive $P - P({\bar P })$ collision.
a) Elastic case, b) Including the diffractive dissociation contribution
\end{document}